\documentclass[10pt,a4paper]{article}
\usepackage{amsmath,amssymb,amsfonts}
\usepackage{graphicx}
\usepackage{hyperref}
\usepackage[margin=0.75in]{geometry}
\usepackage{float}
\usepackage{bm}
\usepackage{booktabs}
\usepackage{tabularx}
\usepackage{subcaption}

\usepackage{setspace}
\usepackage{titlesec}
\titlespacing*{\section}{0pt}{8pt}{4pt}
\titlespacing*{\subsection}{0pt}{6pt}{3pt}

\AtBeginDocument{
  \setlength{\abovedisplayskip}{4pt}
  \setlength{\belowdisplayskip}{4pt}
  \setlength{\abovedisplayshortskip}{2pt}
  \setlength{\belowdisplayshortskip}{2pt}
}

\let\oldbibliography\thebibliography
\renewcommand{\thebibliography}[1]{%
  \oldbibliography{#1}%
  \setlength{\itemsep}{0pt}%
  \setlength{\parskip}{0pt}%
}

\newcommand{\Dc}{{}^{C}\!D}
\newcommand{\Drl}{{}^{RL}\!D}

\title{An Exactly Solvable Ekman Layer with a Fractional-Order Stress
Closure}

\author{
Sandy Hardian Susanto Herho$^{1,2}$,
Rizki Dimas Permana$^{3,4}$,
Iwan Pramesti Anwar$^{4}$,\\
Rusmawan Suwarman$^{5}$,
Deny Juanda Puradimaja$^{1}$, and
Dasapta Erwin Irawan$^{1,*}$
}

\date{}

\begin{document}
\maketitle

\begin{center}
\small
$^{1}$Applied Geology Research Group, Bandung Institute of Technology,
Bandung, West Java 40132, Indonesia\\
$^{2}$Center for Agrarian Studies, Bandung Institute of Technology,
Bandung, West Java 40132, Indonesia\\
$^{3}$Department of Marine Environmental Sciences, Sumatera Institute of Technology,
Southern Lampung, Lampung 35365, Indonesia\\ 
$^{4}$Applied and Environmental Oceanography Research Group, Bandung
Institute of Technology, Bandung, West Java 40132, Indonesia\\
$^{5}$Atmospheric Science Research Group, Bandung Institute of Technology,
Bandung, West Java 40132, Indonesia\\
$^{*}$Correspondence: dasaptaerwin@itb.ac.id
\end{center}

\begin{abstract}
\noindent
The classical theory of the wind-driven surface layer of a rotating ocean
closes the momentum balance with a local flux-gradient law, in which the
turbulent stress at a given depth is proportional to the shear at that
depth. It predicts a surface current deflected forty-five degrees from the
wind, which exceeds most direct measurements. This study asks what follows
when locality is relaxed. Beginning from the exact integral relation between
turbulent stress and mean shear, and requiring the memory kernel to carry no
preferred vertical scale, we obtain a power-law kernel and therefore a
stress law of fractional order. The resulting equation cannot be posed on
the velocity, because the fractional derivative of a bounded profile
vanishes at the surface, so the wind stress cannot be applied, while the
alternative definition of the derivative leaves the surface current
unbounded. Posed on the stress instead, the problem is solvable in closed
form in Mittag-Leffler functions at every order between zero and one. The
surface deflection then depends on the closure order alone and is smaller
than forty-five degrees throughout, whereas the depth-integrated transport
stays exactly normal to the wind, because that constraint follows from the
momentum balance and not from the closure. The far field decays as a power
law rather than exponentially, and its amplitude vanishes in the local
limit, so that limit is singular. Under a suddenly applied stress the
surface transient decays algebraically. Three independent algorithms agree
closely, and no observational or model data are used.
\end{abstract}


\section{Introduction}
\label{sec:intro}

The theory of the wind-driven boundary layer of a rotating ocean balances
the Coriolis acceleration against the divergence of a turbulent stress taken
proportional to the local vertical shear \cite{Ekman1905}. It predicts a
current that spirals clockwise with depth in the Northern Hemisphere, decays
exponentially over a single depth scale, and is deflected $45^{\circ}$ to
the right of the wind at the surface. The associated depth-integrated
transport, directed normal to the wind, has been confirmed by direct
measurement \cite{Price1987, Chereskin1995}. The surface deflection has not.
Reported values cluster well below $45^{\circ}$ and span a wide range across
sites and seasons \cite{Chereskin1995, Lenn2009, Bressan2019}, and observed
spirals are systematically flatter than the classical form, with rotation
depth scales exceeding amplitude decay scales by factors of two to four
\cite{Price1999, Lenn2009}, a result also obtained from spectral analyses of
Southern Ocean drifter and mooring records \cite{Elipot2009}. Part of that discrepancy has been traced to
contamination by depth-varying geostrophic shear, the removal of which
brings Drake Passage and Kerguelen observations back into agreement with a
constant-viscosity model \cite{Polton2013, Roach2015}. Other departures are
harder to absorb. Multi-year records from a moored buoy in the Bay of Bengal
show surface currents deflected to the left of a clockwise-rotating land
breeze in the Northern Hemisphere, a configuration admitted by the
time-dependent theory at superinertial forcing but seldom documented
\cite{McPhaden2024}.

The usual response to these discrepancies retains the local flux-gradient
law and gives the eddy viscosity vertical structure. An eddy viscosity that
varies gradually with depth yields a deflection larger than $45^{\circ}$,
whereas one concentrated near the surface yields a smaller one
\cite{Grisogono1995, Constantin2021, Bressan2019}, and the
piecewise-constant case admits a closed-form solution that resolves the
dependence of the deflection on the profile \cite{Dritschel2020}. Diurnal
cycling of near-surface stratification reproduces the observed flattening
without altering the closure \cite{Price1999}, and generalized models that
add surface waves and horizontal buoyancy gradients extend the same
framework further \cite{Polton2005, McWilliams2012, Wenegrat2016}. A
different concession appears in the boundary-layer schemes used in
circulation models, which add a countergradient term to the diffusive flux
because a purely local diffusivity cannot represent transport by eddies
whose size is comparable with the layer depth \cite{Large1994}. That
correction admits nonlocality while preserving the differential form of the
closure.

A separate literature questions the locality of the flux-gradient relation
itself. The limitations of gradient transport models have been recognized
for decades \cite{Corrsin1974, Berkowicz1980, Stull1984}, and exact
expressions obtained from the Green's function of the fluctuation equation
show the eddy diffusivity to be an integral operator acting over a finite
upstream region rather than a coefficient \cite{Hamba2022, Hamba2023,
Hamba2025}. Macroscopic forcing methods recover the same operator
numerically and have yielded systematic finite-rank approximations to it
\cite{ManiPark2021, Shirian2022, Liu2023}. When the kernel of such an
operator is a power law, the operator is a fractional derivative. This
identification underlies variable-order fractional models of the Reynolds
stress in wall-bounded flow \cite{Song2021}, fractional subgrid-scale
closures for scalar turbulence \cite{AkhavanSafaei2023}, and nonlocal
generalizations of zero-equation models \cite{Egolf2017}. The mathematical
apparatus is long established: a power-law memory kernel produces the
constitutive law introduced by Scott Blair for materials intermediate
between a Hookean solid and a Newtonian fluid \cite{ScottBlair1947,
ScottBlair1949, Caputo1967}, and the associated transport problems are
governed by fractional kinetic equations whose solutions involve
Mittag-Leffler rather than exponential functions \cite{MetzlerKlafter2000,
Podlubny1999, Gorenflo2014}.

These two lines of work have not been combined on the rotating problem. In a
non-rotating layer a fractional closure alters the shape of a monotone
profile. In a rotating layer the stress is complex, and the operator
governing the solution acquires a branch point at the origin whose
contribution competes with the characteristic roots, so the order of the
closure controls the direction of the flow as well as its magnitude. Whether
that competition leaves a tractable problem, and whether the surface
deflection responds to the closure order in a manner consistent with
measurement, cannot be settled in advance.

We therefore derive the fractional closure from the exact integral relation
between stress and shear, impose it on the rotating momentum balance, and
solve the resulting problem. The first result is structural and negative:
the equation cannot be posed on the velocity, because the Caputo derivative
of a bounded profile vanishes at the surface and the Riemann-Liouville
derivative diverges there, so in the first case the wind stress cannot be
applied and in the second the surface current is unbounded. Posed on the
stress, the problem is exactly solvable. We report closed forms for the
surface deflection, the depth scale, the far field, the deep flow direction,
and the response to a suddenly applied stress, together with numerical
verification against algorithms that share no code path. No model output and
no observational or reanalysis data enter at any stage; every quantity
reported follows from the model equations and, where a closed form exists,
is checked against it.

\section{Methods}
\label{sec:methods}

\subsection{Model description}
\label{subsec:model}

Consider a horizontally homogeneous, unstratified ocean of constant density
$\rho$ on an $f$-plane away from the equator, forced at the surface by a
steady wind stress. Let $\zeta = -z \geq 0$ denote depth below the surface
and collect the horizontal velocity and the horizontal turbulent stress into
the complex variables
\begin{equation}
    \psi \equiv u + i v,
    \qquad
    T \equiv T_{x} + i T_{y}.
    \label{eq:complex}
\end{equation}
Reynolds averaging the horizontal momentum equations and discarding the
horizontal derivatives leaves $-\rho f v = \partial_{z}T_{x}$ and
$\rho f u = \partial_{z}T_{y}$, where $T_{i} = -\rho\langle w'u_{i}'\rangle$
is the turbulent momentum flux. Multiplying the second by $i$, adding, and
converting to the depth coordinate gives
\begin{equation}
    i f \rho\, \psi = -\frac{\mathrm{d}T}{\mathrm{d}\zeta},
    \qquad
    T(0) = \tau,
    \qquad
    T(\infty) = 0,
    \label{eq:momentum}
\end{equation}
with $\tau$ the applied surface stress. Integrating Eq.~(\ref{eq:momentum})
across the layer and applying both boundary conditions gives
\begin{equation}
    \int_{0}^{\infty} \psi \, \mathrm{d}\zeta
      = -\frac{i\tau}{\rho f}.
    \label{eq:transport}
\end{equation}
Equation~(\ref{eq:transport}) states that the transport is exactly normal to
the wind, to the right in the Northern Hemisphere. It follows from the
momentum balance and the boundary conditions alone and involves no closure,
so it holds for every member of the family constructed below and serves as
an exact constraint against which the numerics are tested.

Equation~(\ref{eq:momentum}) is not closed, since $T$ is a property of the
turbulence. The relation between the flux and the mean field is not a matter
of choice at the outset. Writing the fluctuation equation for $\psi'$ with
the mean shear as its source and inverting it with the associated Green's
function yields an exact expression for the flux as a linear functional of
the mean gradient \cite{Corrsin1974, Hamba2022},
\begin{equation}
    T(\zeta)
      = -\rho \int_{0}^{\infty}
        D(\zeta,\xi)\,
        \frac{\mathrm{d}\psi}{\mathrm{d}\xi}(\xi)\,
        \mathrm{d}\xi ,
    \label{eq:exactflux}
\end{equation}
in which $D$ is an eddy diffusivity kernel rather than a coefficient. Direct
numerical simulation confirms that $D$ has finite width and that replacing
it by a coefficient can be substantially in error \cite{Hamba2022,
Hamba2023, Hamba2025}, and the same operator has been recovered
independently by macroscopic forcing \cite{ManiPark2021, Shirian2022,
Liu2023}. The local law is the leading term of a Kramers-Moyal expansion of
Eq.~(\ref{eq:exactflux}): expanding the mean shear about $\xi = \zeta$ and
truncating at zeroth order gives
$T \simeq -\rho K \,\mathrm{d}\psi/\mathrm{d}\zeta$ with
$K = \int D\,\mathrm{d}\xi$. That truncation is legitimate only when the
kernel possesses a finite width, that is only when the turbulence has a
mixing length short compared with the scale over which the mean shear
varies.

\begin{figure}[H]
    \centering
    \includegraphics[width=0.52\linewidth]{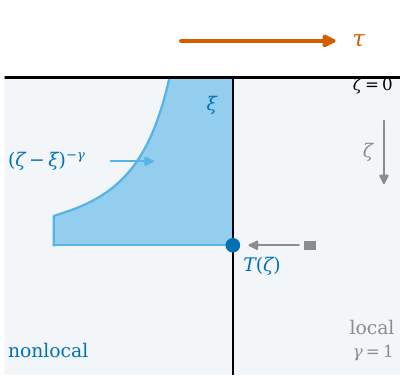}
    \caption{The nonlocal stress closure of Eq.~(\ref{eq:closure}). The
    stress at depth $\zeta$ is a power-law-weighted integral of the shear at
    all depths $\xi < \zeta$ above it, with kernel $(\zeta-\xi)^{-\gamma}$
    drawn here at $\gamma = 0.6$ for legibility. The grey marker on the
    right indicates the local limit $\gamma = 1$, at which the same weight
    collapses onto the evaluation depth and the closure reduces to the
    Fickian law. The wind stress $\tau$ is applied at $\zeta = 0$ and depth
    increases downward.}
    \label{fig:schematic}
\end{figure}

In the wind-driven surface layer that condition fails by construction. The
energy-containing eddies range from the wave and roughness scale up to the
depth of the layer itself, and the mean shear varies over that same depth,
so no scale separation exists and the kernel has no characteristic width.
The natural formalization of this statement is to require the kernel to be
free of any preferred vertical scale, that is to retain its form under a
rescaling of depth,
\begin{equation}
    D(\lambda\zeta, \lambda\xi) = \lambda^{-\gamma} D(\zeta,\xi),
    \qquad \lambda > 0 .
    \label{eq:selfsimilar}
\end{equation}
For a kernel that depends on its arguments only through their difference,
Eq.~(\ref{eq:selfsimilar}) admits the single solution
$D \propto (\zeta-\xi)^{-\gamma}$, since power laws are the only
homogeneous functions of a single variable. Two further physical
requirements fix the domain of the exponent. Integrability of the kernel at
$\xi \to \zeta$ requires $\gamma < 1$, and a flux that vanishes with the
shear requires $\gamma > 0$; the endpoint $\gamma = 1$ recovers a delta
kernel and hence the local law. Finally, momentum enters at the surface and
is carried downward, so the flux at a given depth is set by the shear the
eddies have traversed on the way down. The kernel therefore has support
above the evaluation depth only, and Eq.~(\ref{eq:exactflux}) becomes
\begin{equation}
    T(\zeta)
      = -\frac{\rho K_{\gamma}}{\Gamma(1-\gamma)}
        \int_{0}^{\zeta}
        \frac{1}{(\zeta-\xi)^{\gamma}}\,
        \frac{\mathrm{d}\psi}{\mathrm{d}\xi}(\xi)\,
        \mathrm{d}\xi
      \equiv -\rho K_{\gamma}\, \Dc^{\gamma}_{\zeta}\psi,
    \qquad
    0 < \gamma \leq 1,
    \label{eq:closure}
\end{equation}
where the normalization by $\Gamma(1-\gamma)$ is chosen so that the operator
is exactly the Caputo fractional derivative of order $\gamma$ based at the
surface \cite{Caputo1967, Podlubny1999, Diethelm2010}. The closure
coefficient $K_{\gamma}$ carries units of
$\mathrm{m}^{1+\gamma}\,\mathrm{s}^{-1}$ and reduces to an eddy viscosity at
$\gamma = 1$, where Eq.~(\ref{eq:closure}) is the Fickian law exactly. The
same constitutive form describes materials intermediate between an elastic
solid and a viscous fluid \cite{ScottBlair1947, ScottBlair1949}, and the
divergence of the first moment of the kernel for every $\gamma < 1$ is the
precise statement that no finite mixing length exists and that the local
truncation is unavailable. The structure of the closure is illustrated in
Fig.~\ref{fig:schematic}.

Substituting Eq.~(\ref{eq:closure}) into Eq.~(\ref{eq:momentum}) formally
gives a fractional differential equation of order $\mu \equiv 1+\gamma$ for
$\psi$. That formulation cannot be completed. For any $\psi$ with bounded
derivative near the surface, the integral in Eq.~(\ref{eq:closure}) is
$O(\zeta^{1-\gamma})$ as $\zeta \to 0$, so
$\Dc^{\gamma}_{\zeta}\psi \to 0$ and the closure forces $T(0) = 0$ at every
$\gamma < 1$, which makes the wind stress impossible to impose. Replacing
the Caputo derivative by its Riemann-Liouville counterpart does not help,
since the two differ by
\begin{equation}
    \Drl^{\gamma}_{\zeta}\psi - \Dc^{\gamma}_{\zeta}\psi
      = \frac{\psi(0)}{\Gamma(1-\gamma)}\,\zeta^{-\gamma},
    \label{eq:rlvscaputo}
\end{equation}
which diverges at the surface unless $\psi(0) = 0$. The surface stress is
then either unbounded or the surface current vanishes. The degeneracy is a
property of the operator rather than of any particular solution, and it is
the reason the problem must be posed on the stress.

Applying Eq.~(\ref{eq:closure}) to Eq.~(\ref{eq:momentum}) in the opposite
order eliminates $\psi$ and leaves
\begin{equation}
    \Dc^{\gamma}_{\zeta}\!\left(\frac{\mathrm{d}T}{\mathrm{d}\zeta}\right)
      = b\, T,
    \qquad
    b \equiv \frac{i f}{K_{\gamma}} .
    \label{eq:stressode}
\end{equation}
Let $\hat{T}(p)$ denote the Laplace transform of $T$ in depth. Using
$\mathcal{L}\{\Dc^{\gamma}g\} = p^{\gamma}\hat{g} - p^{\gamma-1}g(0)$ with
$g = \mathrm{d}T/\mathrm{d}\zeta$, so that $\hat{g} = p\hat{T}-\tau$ and
$g(0) = -i f\rho\,\psi_{0}$ with $\psi_{0} \equiv \psi(0)$,
Eq.~(\ref{eq:stressode}) transforms to
\begin{equation}
    \hat{T}(p)
      = \frac{\tau p^{\gamma} - i f \rho \psi_{0}\, p^{\gamma-1}}
             {p^{\mu} - b},
    \qquad
    \mu = 1+\gamma .
    \label{eq:Thatraw}
\end{equation}
The characteristic equation $p^{\mu} = b$ has principal root
\begin{equation}
    p_{0} = b^{1/\mu},
    \qquad
    \arg p_{0} = \frac{\pi}{2\mu}
      \in \left[\frac{\pi}{4},\frac{\pi}{2}\right),
    \label{eq:root}
\end{equation}
which lies in the right half plane for every $\gamma \in (0,1]$ and
therefore carries a mode growing with depth. Imposing $T(\infty)=0$ requires
the numerator of Eq.~(\ref{eq:Thatraw}) to vanish at $p_{0}$, which fixes
the surface velocity and reduces the transform to
\begin{equation}
    \psi_{0} = \frac{\tau p_{0}}{i f \rho},
    \qquad
    \hat{T}(p)
      = \tau\,\frac{p^{\gamma-1}\left(p-p_{0}\right)}{p^{\mu}-b},
    \qquad
    \hat{\psi}(p)
      = -\frac{p\hat{T}(p) - \tau}{i f \rho} .
    \label{eq:That}
\end{equation}
The surface velocity is not prescribed but determined by the far-field
condition, exactly as in the classical problem.

Inverting Eq.~(\ref{eq:That}) term by term with
$\mathcal{L}^{-1}\{p^{\mu-\nu}/(p^{\mu}-b)\}
= \zeta^{\nu-1}E_{\mu,\nu}(b\zeta^{\mu})$, where $E_{\mu,\nu}$ is the
two-parameter Mittag-Leffler function \cite{Gorenflo2014, Mainardi2010},
gives the exact stress profile
\begin{equation}
    T(\zeta)
      = \tau\left[
          E_{\mu,1}\!\left(b\zeta^{\mu}\right)
          - p_{0}\zeta\, E_{\mu,2}\!\left(b\zeta^{\mu}\right)
        \right],
    \label{eq:Tseries}
\end{equation}
and differentiating termwise and using Eq.~(\ref{eq:momentum}) gives the
velocity
\begin{equation}
    \psi(\zeta)
      = -\frac{\tau}{i f \rho}\left[
          b\zeta^{\gamma} E_{\mu,\mu}\!\left(b\zeta^{\mu}\right)
          - p_{0} E_{\mu,1}\!\left(b\zeta^{\mu}\right)
        \right].
    \label{eq:psiseries}
\end{equation}
At $\gamma = 1$ these reduce to $T = \tau e^{-p_{0}\zeta}$ and
$\psi = \psi_{0}e^{-p_{0}\zeta}$ with $p_{0} = \sqrt{if/K}$, the classical
Ekman spiral. The first term of Eq.~(\ref{eq:psiseries}) carries
$\zeta^{\gamma}$ and vanishes at the surface, which is consistent with
$\psi(0) = \tau p_{0}/(if\rho)$ and is the analytical counterpart of the
degeneracy identified above.

Several quantities follow immediately. Writing
$\theta \equiv \arg\psi(0) - \arg\tau$ for the surface deflection measured
from the wind direction, Eqs.~(\ref{eq:That}) and (\ref{eq:root}) give
\begin{equation}
    \theta = \frac{\pi}{2\mu} - \frac{\pi}{2}
           = -\frac{\pi}{2}\,\frac{\gamma}{1+\gamma},
    \label{eq:theta}
\end{equation}
which is $-45^{\circ}$ at $\gamma = 1$ and decreases in magnitude
monotonically as the closure becomes more nonlocal. The corresponding depth
scale is the generalized Ekman depth
\begin{equation}
    \delta_{\gamma}
      = \left(\frac{K_{\gamma}}{f}\right)^{1/(1+\gamma)},
    \label{eq:delta}
\end{equation}
reducing to $\sqrt{K/f}$ at $\gamma = 1$. As $p \to 0$ the transform behaves
as $\hat{T} \sim \tau(p_{0}/b)p^{\gamma-1}$, whose inverse is algebraic, so
the far field is
\begin{equation}
    T \sim A_{\gamma}\zeta^{-\gamma},
    \qquad
    \psi \sim \frac{\gamma A_{\gamma}}{i f\rho}\,\zeta^{-(1+\gamma)},
    \qquad
    A_{\gamma} = \frac{\tau\left(p_{0}/b\right)}{\Gamma(1-\gamma)} .
    \label{eq:tail}
\end{equation}
The decay is a power law rather than an exponential, and the amplitude
carries the factor $1/\Gamma(1-\gamma)$, which vanishes identically at
$\gamma = 1$. That factor is the mechanism by which the branch-cut
contribution disappears in the local limit, and it makes the limit singular
rather than continuous. The argument of the far field is independent of
depth,
\begin{equation}
    \arg\psi(\infty) - \arg\tau
      = \frac{\pi}{2\mu} - \pi,
    \label{eq:asymptotic}
\end{equation}
so subtracting Eq.~(\ref{eq:theta}) yields the identity
\begin{equation}
    \arg\psi(\infty) - \arg\psi(0)
      = -\frac{\pi}{2}
      \qquad \left(\mathrm{mod}\ 2\pi\right),
    \label{eq:quarterturn}
\end{equation}
valid at every $\gamma \in (0,1)$. The dependence on $\gamma$ cancels
exactly, so the deep flow direction is slaved to the surface direction
throughout the family. This has no classical counterpart, since at
$\gamma = 1$ the spiral winds without bound and possesses no asymptotic
direction.

Equation~(\ref{eq:quarterturn}) constrains the deep direction and not the
number of revolutions taken to reach it. Writing $n(\gamma)$ for that
winding number, the net turning is $-\pi/2 - 2\pi n$, and $n$ is not
identically zero. Its origin lies in the root structure of $p^{\mu} = b$.
The roots are
$p_{k} = |b|^{1/\mu}\exp\!\left[i(\pi/2 + 2\pi k)/\mu\right]$, and the
$k = -1$ root satisfies $|\arg p_{-1}| = (3\pi/2)/\mu < \pi$ precisely when
$\gamma > 1/2$. Below that order the growing mode has been cancelled and no
pole remains on the principal sheet, so the entire solution is the
branch-cut contribution. Above it a pole is present at $p_{-1}$, with
$\mathrm{Re}\,p_{-1} < 0$, contributing a decaying oscillation of residue
\begin{equation}
    R_{\gamma} = -\frac{\tau\left(p_{-1}-p_{0}\right)}{\mu\, i f \rho}
    \label{eq:residue}
\end{equation}
to the velocity. The velocity vector rotates while that contribution
dominates and ceases to rotate once the algebraic tail of
Eq.~(\ref{eq:tail}) takes over. Equating the two magnitudes defines a
crossover depth $\zeta_{c}$ through
\begin{equation}
    \left|R_{\gamma}\right| e^{\mathrm{Re}(p_{-1})\zeta_{c}}
      = \frac{\gamma\left|A_{\gamma}\right|}{f\rho}\,
        \zeta_{c}^{-(1+\gamma)},
    \label{eq:crossover}
\end{equation}
and predicts a winding number of order
$\left|\mathrm{Im}\,p_{-1}\right|\zeta_{c}/2\pi$. Since
$1/\Gamma(1-\gamma) \approx (1-\gamma)$ as $\gamma \to 1$, and since
$\left|\mathrm{Im}\,p_{-1}/\mathrm{Re}\,p_{-1}\right| \to 1$ in the same
limit, Eq.~(\ref{eq:crossover}) gives
\begin{equation}
    n(\gamma) \sim \frac{1}{2\pi}\ln\!\frac{1}{1-\gamma}
      + \mathcal{O}(1),
    \qquad
    \gamma \to 1^{-} .
    \label{eq:windinglaw}
\end{equation}
This is an estimate rather than a theorem, since it neglects phase relations
and prefactors of order unity. It is reported because it identifies the
mechanism and predicts a logarithmic rather than an algebraic divergence,
and its accuracy is assessed against measured winding numbers below. The
same argument accounts for the absence of any signature at $\gamma = 1/2$,
where $p_{-1}$ enters the principal sheet: at that order the tail amplitude
is of order unity, so the pole is subdominant at every depth and cannot
produce a complete revolution. The crossing is necessary for winding but not
sufficient.

The response to a suddenly applied stress follows from restoring the
tendency term,
\begin{equation}
    \rho\left(\frac{\partial \psi}{\partial t} + i f \psi\right)
      = -\frac{\partial T}{\partial \zeta},
    \qquad
    T(0,t) = \tau H(t),
    \qquad
    \psi(\zeta,0) = 0,
    \label{eq:unsteady}
\end{equation}
with $H$ the Heaviside function. Transforming in time replaces $if$ by
$\sigma + if$ throughout, so the depth structure derived above carries over
with $b_{\sigma} = (\sigma+if)/K_{\gamma}$ and
$p_{\sigma} = b_{\sigma}^{1/\mu}$. At the surface this collapses to a single
elementary transform,
\begin{equation}
    \tilde{\psi}(0,\sigma)
      = \frac{\tau}{\rho}\,K_{\gamma}^{-1/\mu}\,
        \frac{\left(\sigma + i f\right)^{-a}}{\sigma},
    \qquad
    a \equiv \frac{\gamma}{1+\gamma},
    \label{eq:spinuptransform}
\end{equation}
whose inverse is exact. Convolving
$\mathcal{L}^{-1}\{(\sigma+if)^{-a}\} = e^{-ift}t^{a-1}/\Gamma(a)$ with the
step and normalizing by the steady value gives
\begin{equation}
    \psi(0,t) = \psi(0,\infty)\, P\!\left(a,\, i f t\right),
    \qquad
    P(a,z) = \frac{1}{\Gamma(a)}\int_{0}^{z} e^{-s}s^{a-1}\,\mathrm{d}s,
    \label{eq:spinup}
\end{equation}
with $P$ the regularized lower incomplete gamma function. The exponent $a$
is the same combination of $\gamma$ that fixes the deflection, since
Eq.~(\ref{eq:theta}) reads $\theta = -\tfrac{\pi}{2}a$. At $\gamma = 1$ one
has $a = 1/2$ and $P(1/2,z) = \mathrm{erf}(\sqrt{z})$, recovering the
classical impulsively started surface current. For $\gamma < 1$ the
asymptotic form $\Gamma(a,z) \sim z^{a-1}e^{-z}$ gives a residual inertial
oscillation whose envelope decays as
\begin{equation}
    \left|1 - \frac{\psi(0,t)}{\psi(0,\infty)}\right|
      \sim \frac{(ft)^{a-1}}{\Gamma(a)},
    \qquad
    a - 1 = -\frac{1}{1+\gamma},
    \label{eq:envelope}
\end{equation}
so the surface transient is forgotten algebraically at every $\gamma$,
including the classical value, and never exponentially.

An equivalent formulation without derivatives is obtained by applying the
fractional integral $I^{\gamma}$ to Eq.~(\ref{eq:closure}), which gives
$I^{\gamma}\Dc^{\gamma}\psi = \psi - \psi_{0}$, and integrating
Eq.~(\ref{eq:momentum}) directly:
\begin{equation}
    \psi(\zeta) = \psi_{0} - \frac{1}{\rho K_{\gamma}} I^{\gamma}T(\zeta),
    \qquad
    T(\zeta) = \tau - i f \rho \int_{0}^{\zeta}\psi(\xi)\,\mathrm{d}\xi .
    \label{eq:volterra}
\end{equation}
This is a coupled weakly singular Volterra system of the second kind and
provides a route to the solution that never touches the Laplace plane.

It remains to establish how many of the dimensional parameters actually
control the solution, since every quantity reported below is dimensionless
and the choice of units is not merely presentational. The problem is
specified by $f$, $K_{\gamma}$,
$\rho$, and $\tau$, together with the dimensionless order $\gamma$, and the
purpose of scaling is to establish how many of these actually control the
solution.

The system of Eqs.~(\ref{eq:momentum}) and (\ref{eq:closure}) is linear and
homogeneous in $\tau$, so the surface stress enters only as a multiplicative
amplitude and can be set to unity without loss of generality. The three
remaining parameters have dimensions
$[f] = \mathrm{T}^{-1}$,
$[K_{\gamma}] = \mathrm{L}^{1+\gamma}\mathrm{T}^{-1}$, and
$[\rho] = \mathrm{M}\mathrm{L}^{-3}$. Seeking a dimensionless combination
$f^{\alpha}K_{\gamma}^{\beta}\rho^{c}$ gives $c = 0$ from the mass
dimension, then $\beta = 0$ from the length dimension, then $\alpha = 0$
from the time dimension. No nontrivial group exists. The three parameters
therefore fix the mass, length, and time units uniquely, through $\rho$, the
depth scale $\delta_{\gamma}$ of Eq.~(\ref{eq:delta}), and the inertial time
$1/f$, and the reduced problem depends on $\gamma$ alone. Scaling $\zeta$ by
$\delta_{\gamma}$, $\psi$ by
$|\tau|\delta_{\gamma}^{\gamma}/(\rho K_{\gamma})$, $t$ by $1/f$, and $T$ by
$|\tau|$ realizes this reduction, and the reference case
$f = K_{\gamma} = \rho = \tau = 1$ sets $\delta_{\gamma} = 1$ by
construction. Three consequences follow. The entire one-parameter family can
be mapped by varying a single number. Results obtained at one set of
physical scales transfer to every other set. And any dependence on
$f$, $K_{\gamma}$, or $\rho$ that appears in a computed quantity is a
numerical artifact rather than physics.

One caveat limits what may legitimately be compared across the family. Since
$K_{\gamma}$ carries units of
$\mathrm{m}^{1+\gamma}\mathrm{s}^{-1}$, it is a dimensionally different
quantity at each $\gamma$, and setting $K_{\gamma} = 1$ therefore does not
fix a common physical mixing strength. The absolute depth scale
$\delta_{\gamma}$ in metres, the absolute surface speed, and the absolute
stress at a given depth in metres are consequently not comparable between
different $\gamma$ at fixed numerical $K_{\gamma}$, and none is reported
here as a physical number. What is comparable, because it is independent of
$K_{\gamma}$ entirely, is the surface deflection of Eq.~(\ref{eq:theta}),
the deep direction of Eq.~(\ref{eq:asymptotic}), the turning of
Eq.~(\ref{eq:quarterturn}), the winding number, the decay exponent
$1+\gamma$, the spin-up exponent $a$, and every ratio plotted against
$\zeta/\delta_{\gamma}$. Those are the quantities this study reports. For
the same reason the closure kernel is plotted only in the normalized form
$\zeta^{\gamma}(\zeta-\xi)^{-\gamma}/\Gamma(1-\gamma)$, which removes its
$\mathrm{m}^{-\gamma}$ units.

\subsection{Numerical implementation}
\label{subsec:numerics}

The closed forms derived above are exact, but their evaluation is not
trivial: the Mittag-Leffler series loses relative
precision through subtractive cancellation at large argument, and contour
methods lose it at small depth. Three algorithms sharing no code path were
therefore implemented, so that agreement between them tests the formulation
rather than testing one implementation against itself. The protocol follows
the practice adopted in related idealized solvers, in which correctness is
established against closed-form solutions, conserved quantities, and measured
convergence orders rather than against observations \cite{Herho2025kh2d,
Irawan2026amerta, Irawan2026kdv, Herho2026wave}.

The first algorithm inverts Eq.~(\ref{eq:That}) numerically. Starting from
the Bromwich integral
$T(\zeta) = (2\pi i)^{-1}\int e^{p\zeta}\hat{T}(p)\,\mathrm{d}p$ and
deforming the contour onto the cotangent curve
\begin{equation}
    p(\vartheta) = r\vartheta\left(\cot\vartheta + i\right),
    \qquad
    \vartheta \in (-\pi,\pi),
    \qquad
    r = \frac{2M}{5\zeta},
    \label{eq:contour}
\end{equation}
one has
$\mathrm{d}p = r\left(\cot\vartheta - \vartheta\csc^{2}\vartheta + i\right)
\mathrm{d}\vartheta$, and therefore
\begin{equation}
    \frac{\mathrm{d}p}{2\pi i}
      = \frac{r}{2\pi}
        \left[1 + i\left(\vartheta\csc^{2}\vartheta
        - \cot\vartheta\right)\right]\mathrm{d}\vartheta
      = \frac{r}{2\pi}\left[1 + i\sigma(\vartheta)\right]
        \mathrm{d}\vartheta,
    \label{eq:talbotweight}
\end{equation}
where the second equality uses the identity
$\vartheta\csc^{2}\vartheta - \cot\vartheta
= \vartheta + (\vartheta\cot\vartheta - 1)\cot\vartheta
\equiv \sigma(\vartheta)$, which was verified numerically to
$2.2\times10^{-14}$ over the contour. Applying the trapezoidal rule at
$\vartheta_{k} = k\pi/M$ gives the fixed-Talbot estimate
\begin{equation}
    T(\zeta) \simeq \frac{r}{M}\left\{
      \frac{1}{2}e^{r\zeta}\hat{T}(r)
      + \frac{1}{2}\sum_{k=1}^{M-1}\left[
        e^{p_{k}\zeta}\hat{T}(p_{k})w_{k}
        + e^{\bar{p}_{k}\zeta}\hat{T}(\bar{p}_{k})\bar{w}_{k}
      \right]\right\},
    \qquad
    w_{k} = 1 + i\sigma(\vartheta_{k}),
    \label{eq:talbot}
\end{equation}
with $p_{k} = p(\vartheta_{k})$ \cite{Talbot1979, AbateValko2004}. The
contour wraps the negative real axis and therefore handles the branch point
of $p^{\mu}$ at the origin and the cut along the negative axis with no
special treatment, which is what is needed here, since for $\gamma < 1/2$
the transform has no poles at all once the growing mode has been cancelled
and the entire solution is the branch-cut contribution. Standard
implementations halve the work by retaining only the real part of the sum,
which presumes the inverse to be real valued and hence
$\hat{T}(\bar{p}) = \overline{\hat{T}(p)}$. The stress and velocity here are
complex valued functions of a real depth and carry no such symmetry, so
Eq.~(\ref{eq:talbot}) retains both halves of the contour explicitly and
imposes no symmetry. Convergence is geometric in $M$ until roundoff
amplified by $e^{2M/5}$ takes over.

The second algorithm evaluates Eqs.~(\ref{eq:Tseries}) and
(\ref{eq:psiseries}) by summing the defining series
$E_{\mu,\nu}(z) = \sum_{k\geq0} z^{k}/\Gamma(\mu k + \nu)$ at a working
precision chosen adaptively from $|b\zeta^{\mu}|$, using the arbitrary
precision arithmetic of \texttt{mpmath}. The series converges for every
argument, so this route is exact in principle and accurate in practice at
small and moderate depth; it becomes prohibitive at large argument, where
contour methods designed for the Mittag-Leffler function itself are
preferable \cite{Garrappa2015, Gorenflo2002}. The inversion pair on which
Eqs.~(\ref{eq:Tseries}) and (\ref{eq:psiseries}) rest was verified
independently by comparing
$\mathcal{L}^{-1}\{p^{\mu-\nu}/(p^{\mu}-b)\}$ computed by
Eq.~(\ref{eq:talbot}) with
$\zeta^{\nu-1}E_{\mu,\nu}(b\zeta^{\mu})$ computed by series, which agree to
between $2.4\times10^{-12}$ and $1.4\times10^{-11}$ for
$(\mu,\nu) \in \{(1.6,1),(1.6,2),(1.7,1.7)\}$. Because the two routes are
exact in complementary regions, the public solution splices them, using the
series below $\zeta = 0.05$ and the contour above, with the splice point
inside the region where both are accurate to better than $10^{-12}$.
Cross-method agreement is reported on the contour branch alone, so the
splice is never tested against itself.

The third algorithm solves Eq.~(\ref{eq:volterra}) directly. On a uniform
grid $\zeta_{n} = nh$, the fractional integral is
\begin{equation}
    I^{\gamma}T(\zeta_{n})
      = \frac{1}{\Gamma(\gamma)}
        \int_{0}^{\zeta_{n}}
        \left(\zeta_{n}-s\right)^{\gamma-1}T(s)\,\mathrm{d}s .
    \label{eq:fracint}
\end{equation}
Replacing $T$ by its piecewise-linear interpolant on the grid and
integrating the resulting products of the singular weight with the hat
functions analytically gives the product trapezoidal rule
\begin{equation}
    I^{\gamma}T(\zeta_{n})
      \simeq \frac{h^{\gamma}}{\Gamma(\gamma+2)}
        \sum_{j=0}^{n} a_{j,n}T_{j},
    \qquad
    a_{j,n} =
    \begin{cases}
      (n-1)^{\gamma+1} - n^{\gamma}\left(n-\gamma-1\right), & j = 0,\\[2pt]
      (n-j+1)^{\gamma+1} + (n-j-1)^{\gamma+1} - 2(n-j)^{\gamma+1},
        & 1 \leq j \leq n-1,\\[2pt]
      1, & j = n,
    \end{cases}
    \label{eq:piweights}
\end{equation}
which is the standard Diethelm quadrature \cite{DiethelmFordFreed2002,
DiethelmFordFreed2004, Diethelm2010, Lubich1985}. The weights were checked
against the exact value $I^{\gamma}\zeta^{2} = 2\zeta^{2+\gamma}/
\Gamma(3+\gamma)$ at $\zeta = 1$: the error falls from
$4.3\times10^{-6}$ at $N = 200$ to $2.7\times10^{-7}$ at $N = 800$ at
$\gamma = 0.3$, $0.6$, and $0.9$, confirming second-order convergence of the
quadrature itself.

Discretizing the second member of Eq.~(\ref{eq:volterra}) with the
trapezoidal rule and writing $Q_{n} = \int_{0}^{\zeta_{n}}\psi$, the update
at step $n$ is obtained by substituting
\begin{equation}
    T_{n} = \tau - i f \rho\left[
      Q_{n-1} + \tfrac{1}{2}h\left(\psi_{n-1}+\psi_{n}\right)\right]
    \label{eq:Tupdate}
\end{equation}
into the first member, which yields a single linear equation for $\psi_{n}$,
\begin{equation}
    \left(1 - \frac{i f h\, w}{2K_{\gamma}}\right)\psi_{n}
      = \psi_{0}
        - \frac{w}{\rho K_{\gamma}}
          \left[\sum_{j=0}^{n-1}a_{j,n}T_{j}
          + \tau - i f \rho\left(Q_{n-1}
            + \tfrac{1}{2}h\psi_{n-1}\right)\right],
    \qquad
    w = \frac{h^{\gamma}}{\Gamma(\gamma+2)} .
    \label{eq:pistep}
\end{equation}
The step is therefore implicit and solved exactly rather than by a
predictor-corrector iteration. The scheme never evaluates a Mittag-Leffler
function and never enters the Laplace plane, so it is genuinely independent
of the first two algorithms. Its expected order is $\min(2, 1+\gamma)$,
limited not by the quadrature of Eq.~(\ref{eq:piweights}) but by the
$\zeta^{\gamma}$ cusp that Eq.~(\ref{eq:psiseries}) carries at the surface,
which is the standard situation for weakly singular Volterra equations with
nonsmooth solutions \cite{DiethelmFordFreedLuchko2005, Lubich1986,
Dixon1985}. The observed order is itself a diagnostic, since a formulation
error would not reproduce the predicted exponent as a function of $\gamma$.

Seven checks were applied. The surface deflection from the series solution
at $\zeta = 10^{-200}$ was compared with Eq.~(\ref{eq:theta}). The contour
and series velocities were compared over $0.25 < \zeta < 16$. The
compensated speed $\zeta^{1+\gamma}|\psi|$ at $\zeta = 10^{5}$ was compared
with the closed-form amplitude of Eq.~(\ref{eq:tail}) with no fitted
constant. The solution was substituted back into the closure in the integral
form of Eq.~(\ref{eq:volterra}), with $I^{\gamma}T$ evaluated by
arbitrary-precision quadrature of the contour stress rather than by the
product rule, so that the closure itself is tested and not merely the
transform algebra. The transport of Eq.~(\ref{eq:transport}) was evaluated
along both the contour and the product-integration solutions. Finally the
spin-up of Eq.~(\ref{eq:spinup}) was compared both with a nested contour
inversion of Eq.~(\ref{eq:spinuptransform}) and with a direct quadrature of
the convolution, in which the substitution $s = t w^{1/a}$ removes the
endpoint singularity at $s = 0$ analytically, since
$s^{a-1}\mathrm{d}s = (t^{a}/a)\,\mathrm{d}w$. The contour comparison for
the spin-up is valid only while the Talbot contour encloses the branch point
at $\sigma = -if$, that is for $ft < 2\pi M/5$, whereas the quadrature
comparison is valid at any time.

One limitation of the double precision pipeline should be recorded, since it
bears directly on the winding number. As $\gamma \to 1$ the tail amplitude
of Eq.~(\ref{eq:tail}) is proportional to $1-\gamma$ and the velocity decays
as $\zeta^{-(1+\gamma)}$, so at $\gamma = 0.999$ the algebraic contribution
at $\zeta = 10^{6}$ lies roughly fifteen orders of magnitude below the
surface value and is lost in the roundoff of the Talbot sum. The turning
accumulated in double precision therefore ceases to converge beyond
$\zeta \approx 10^{5}$ at the largest orders and drifts rather than
settling. Equation~(\ref{eq:quarterturn}) was accordingly reverified at
$\gamma = 0.999$ in sixty-digit arithmetic with $M = 64$, as reported below.
The double precision residuals at $\zeta = 10^{4}$ remain valid; the
statement that they shrink monotonically applies to exact arithmetic and not
to the double precision evaluation at greater depth.

All computations were carried out in Python. Array operations and the
special functions of the closed forms use \texttt{NumPy} \cite{Harris2020}
and \texttt{SciPy} \cite{Virtanen2020}, arbitrary precision arithmetic uses
\texttt{mpmath}, and figures were produced with \texttt{Matplotlib}
\cite{Hunter2007} using a colorblind-accessible palette \cite{Wong2011}. The
solver is organized as a package in which \texttt{core} holds the parameters
and closed forms, \texttt{transform} the contour inversion,
\texttt{mittagleffler} the arbitrary precision series, \texttt{solution} the
spliced public interface, \texttt{volterra} the product-integration march,
and \texttt{unsteady} the response to a suddenly applied stress. One script
per figure regenerates every number reported.

\section{Results}
\label{sec:results}

\begin{figure}[H]
    \centering
    \includegraphics[width=0.68\linewidth]{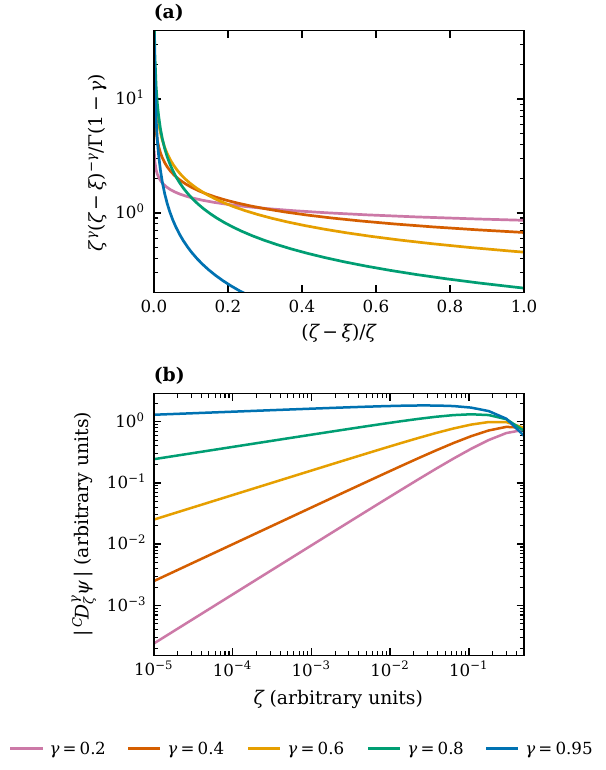}
    \caption{The closure kernel and the surface degeneracy. (a) Normalized
    kernel $\zeta^{\gamma}(\zeta-\xi)^{-\gamma}/\Gamma(1-\gamma)$ against
    the normalized separation $(\zeta-\xi)/\zeta$, at five closure orders.
    Small $\gamma$ weights the entire overlying column comparably; large
    $\gamma$ concentrates the weight at the evaluation depth. (b) Magnitude
    of the Caputo derivative of a bounded test profile against depth,
    showing the $\zeta^{1-\gamma}$ approach to zero that forces $T(0) = 0$
    and prevents the problem from being posed on $\psi$.}
    \label{fig:closure}
\end{figure}

The closure kernel and the degeneracy it induces are shown in
Fig.~\ref{fig:closure}. The normalized kernel is nearly flat at
$\gamma = 0.2$, so that the shear throughout the overlying column
contributes comparably to the stress, and concentrates sharply near the
evaluation depth at $\gamma = 0.95$. The magnitude of the Caputo derivative
of a synthetic bounded test profile tends to zero at the surface at every
$\gamma < 1$, with fitted logarithmic slopes of $0.800$, $0.600$,
$0.400$, $0.199$, and $0.0494$ at $\gamma = 0.2$, $0.4$, $0.6$,
$0.8$, and $0.95$, against the predicted values $1-\gamma$ of $0.8$, $0.6$,
$0.4$, $0.2$, and $0.05$. The residuals range from $3.77 \times 10^{-4}$ to
$6.46 \times 10^{-4}$ and reflect the next-order term of the expansion
rather than a failure of the scaling. The abscissa of the lower panel is in
the arbitrary units of the test profile, so only the slope carries meaning.

\begin{figure}[H]
    \centering
    \includegraphics[width=0.7\linewidth]{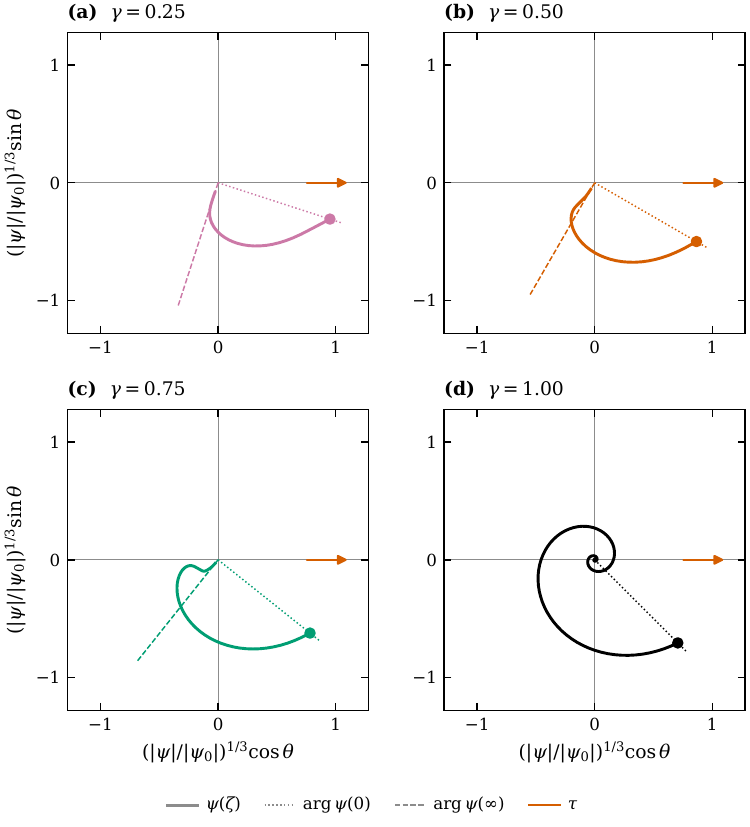}
    \caption{Hodographs of the exact solution at (a) $\gamma = 0.25$,
    (b) $\gamma = 0.50$, (c) $\gamma = 0.75$, and (d) $\gamma = 1.00$. The
    radial coordinate is compressed as $|\psi/\psi_{0}|^{1/3}$ and angles
    are undistorted. Dotted and dashed rays mark $\arg\psi(0)$ and
    $\arg\psi(\infty)$ from Eqs.~(\ref{eq:theta}) and
    (\ref{eq:asymptotic}); the arrow marks the wind stress. The winding
    number is zero at all four orders shown.}
    \label{fig:spiral}
\end{figure}

Hodographs of the exact solution appear in Fig.~\ref{fig:spiral}, with the
radial coordinate compressed as $|\psi/\psi_{0}|^{1/3}$ so that the
classical winding remains visible; the map leaves angles undistorted. The
surface deflection measured from the solution matches
Eq.~(\ref{eq:theta}) at every order tested, giving $-18.0^{\circ}$,
$-30.0^{\circ}$, and $-38.6^{\circ}$ at $\gamma = 0.25$, $0.50$,
and $0.75$, against $-45^{\circ}$ at $\gamma = 1$. The turning modulo
$360^{\circ}$ is $-90^{\circ}$ in all three fractional cases. The
qualitative difference from the classical spiral is apparent: for
$\gamma < 1$ the curve terminates on a fixed asymptotic direction, whereas
at $\gamma = 1$ it spirals into the origin without limit.

Closed-form angles are listed in Table~\ref{tab:angles}. The surface
deflection increases in magnitude monotonically from $-8.18^{\circ}$ at
$\gamma = 0.1$ to $-45^{\circ}$ at $\gamma = 1$, and the deep asymptotic
direction tracks it exactly one quarter turn behind, so the turning modulo
$360^{\circ}$ is $-90^{\circ}$ at every fractional order. The deflection
measured numerically from the series solution agrees with
Eq.~(\ref{eq:theta}) to $1.07\times10^{-14}$ degrees.

\begin{table}[H]
\centering
\caption{Closed-form angles and exponents of the nonlocal Ekman layer from
Eqs.~(\ref{eq:theta}), (\ref{eq:asymptotic}), (\ref{eq:quarterturn}), and
(\ref{eq:spinup}). Angles are in degrees, measured from the wind stress and
negative to the right.}
\label{tab:angles}
\small
\begin{tabular}{ccccc}
\toprule
Order $\gamma$ & Deflection $\theta$ & $\arg\psi(\infty)$ &
Turning (mod $360^{\circ}$) & Exponent $a = \gamma/(1+\gamma)$ \\
\midrule
0.10 & $-8.18$  & $-98.2$  & $-90.0$ & 0.0909 \\
0.20 & $-15.0$  & $-105$   & $-90.0$ & 0.167 \\
0.30 & $-20.8$  & $-111$   & $-90.0$ & 0.231 \\
0.40 & $-25.7$  & $-116$   & $-90.0$ & 0.286 \\
0.50 & $-30.0$  & $-120$   & $-90.0$ & 0.333 \\
0.60 & $-33.8$  & $-124$   & $-90.0$ & 0.375 \\
0.70 & $-37.1$  & $-127$   & $-90.0$ & 0.412 \\
0.80 & $-40.0$  & $-130$   & $-90.0$ & 0.444 \\
0.90 & $-42.6$  & $-133$   & $-90.0$ & 0.474 \\
1.00 & $-45.0$  & none     & unbounded & 0.500 \\
\bottomrule
\end{tabular}
\end{table}

\begin{figure}[H]
    \centering
    \includegraphics[width=0.7\linewidth]{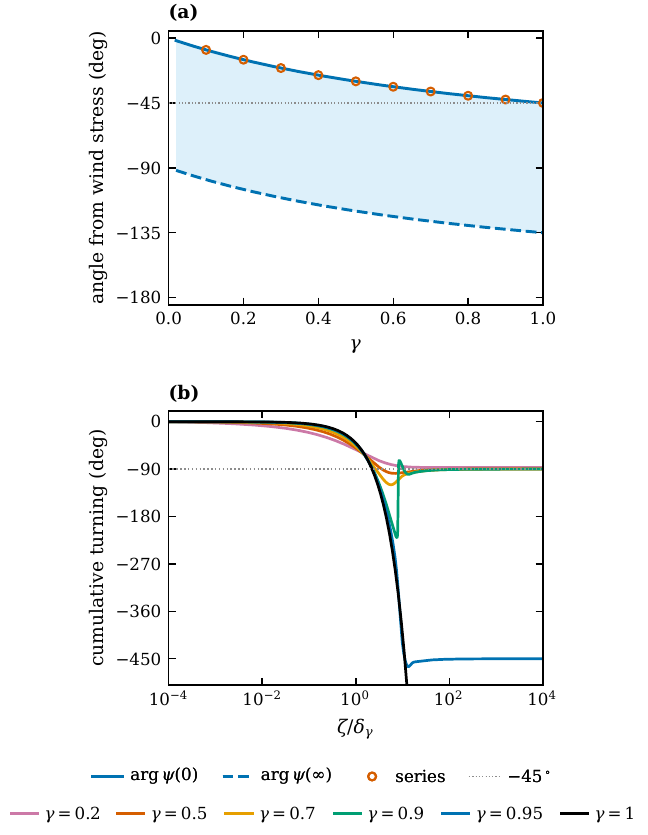}
    \caption{Angles, turning, and winding. (a) Surface deflection and deep
    asymptotic direction against closure order, with the closed forms of
    Eqs.~(\ref{eq:theta}) and (\ref{eq:asymptotic}) drawn as curves and the
    series solution as circles; the shaded band is the quarter turn that
    separates them at every $\gamma$. (b) Cumulative turning of the velocity
    vector against scaled depth at six closure orders. The $\gamma = 0.95$
    curve settles on $-450^{\circ}$ rather than $-90^{\circ}$, and the
    classical curve does not settle at all.}
    \label{fig:angles}
\end{figure}

The angles, the accumulated turning, and the winding number are shown
together in Fig.~\ref{fig:angles}. The measured surface deflection lies on
the closed form throughout. The cumulative turning approaches $-90^{\circ}$
from below at small $\gamma$ and develops an increasingly large excursion
beyond that value as $\gamma$ rises, measuring $3.0^{\circ}$ at
$\gamma = 0.4$, $8.2^{\circ}$ at $\gamma = 0.5$, $29.8^{\circ}$ at
$\gamma = 0.7$, and $130^{\circ}$ at $\gamma = 0.9$. The overshoot grows
smoothly through $\gamma = 1/2$, where the second root of $p^{\mu} = b$
enters the principal sheet, and shows no discontinuity in value or in slope
there. Once the overshoot exceeds $360^{\circ}$ the velocity vector has
completed an additional revolution and the net turning steps from
$-90^{\circ}$ to $-450^{\circ}$; this occurs between $\gamma = 0.90$ and
$\gamma = 0.91$. The narrow spike on the $\gamma = 0.9$ curve near
$\zeta = 7$ is a genuine near-zero of $\psi$, at which the argument is
poorly conditioned, and is not a numerical artifact.

Measured winding numbers are listed in Table~\ref{tab:winding} together with
the crossover estimate of Eq.~(\ref{eq:crossover}). The winding number is
zero up to $\gamma \simeq 0.90$, unity from $\gamma \simeq 0.91$ through
$\gamma = 0.995$, and at least two by $\gamma = 0.999$. The estimate tracks
the measured integer with an approximately constant offset near $0.8$,
rising through unity between $\gamma = 0.90$ and $\gamma = 0.95$ and through
two near $\gamma = 0.999$, which is the behaviour Eq.~(\ref{eq:windinglaw})
predicts. The residual from $-90 - 360n$ degrees is set by truncation at
finite depth rather than by the identity, and falls below $0.01^{\circ}$
wherever the algebraic tail has taken over. At $\gamma = 0.999$ the double
precision residual at $\zeta = 10^{4}$ is $0.107^{\circ}$ and does not
improve with deeper truncation, for the reason given above. Recomputing the
same quantity in sixty-digit arithmetic gives a net turning whose departure
from $-810^{\circ}$ falls from $0.868^{\circ}$ at $\zeta = 10^{2}$ to
$0.008^{\circ}$ at $\zeta = 10^{4}$, to $10^{-4}$ degrees at
$\zeta = 10^{6}$, and below that at $\zeta = 10^{7}$ and beyond, confirming
both $n = 2$ and Eq.~(\ref{eq:quarterturn}) to four decimal places.

\begin{table}[H]
\centering
\caption{Measured winding number of the velocity vector against closure
order, with the net turning evaluated to $\zeta = 10^{4}$, and the crossover
estimate $\left|\mathrm{Im}\,p_{-1}\right|\zeta_{c}/2\pi$ obtained from
Eq.~(\ref{eq:crossover}). No pole is present on the principal sheet for
$\gamma \leq 1/2$.}
\label{tab:winding}
\small
\begin{tabular}{ccccc}
\toprule
Order $\gamma$ & Net turning (deg) & Winding $n$ &
Residual from $-90-360n$ & Crossover estimate \\
\midrule
0.200 & $-86.9$ & 0 & $+3.06$   & --- \\
0.400 & $-89.3$ & 0 & $+0.715$  & --- \\
0.600 & $-89.9$ & 0 & $+0.135$  & 0.110 \\
0.800 & $-90.0$ & 0 & $+0.0171$ & 0.458 \\
0.900 & $-90.0$ & 0 & $+0.0026$ & 0.790 \\
0.910 & $-450$  & 1 & $+0.0018$ & 0.835 \\
0.950 & $-450$  & 1 & $-0.0002$ & 1.07 \\
0.990 & $-450$  & 1 & $+0.0082$ & 1.53 \\
0.995 & $-450$  & 1 & $+0.0193$ & 1.69 \\
0.999 & $-810$  & 2 & $+0.107$  & 2.01 \\
\bottomrule
\end{tabular}
\end{table}

\begin{figure}[H]
    \centering
    \includegraphics[width=\linewidth]{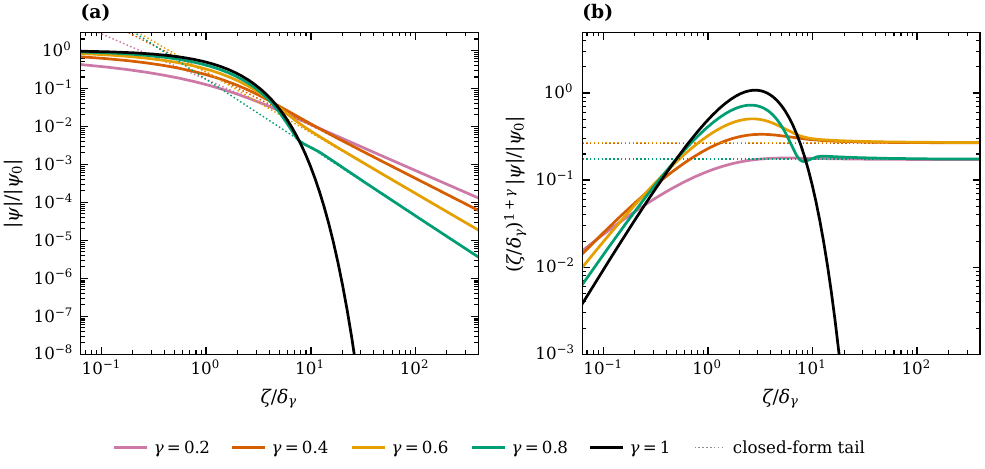}
    \caption{Depth profiles and the algebraic far field. (a) Normalized
    speed against scaled depth at five closure orders, with the closed-form
    tail of Eq.~(\ref{eq:tail}) dotted. (b) The same profiles compensated by
    $(\zeta/\delta_{\gamma})^{1+\gamma}$, which collapses each fractional
    case onto the closed-form amplitude
    $\gamma|A_{\gamma}|/(f\rho)$. At $\gamma = 1$ the amplitude is
    identically zero and the compensated profile falls without limit.}
    \label{fig:profiles}
\end{figure}

Depth profiles and the algebraic far field are shown in
Fig.~\ref{fig:profiles}. Speeds are normalized by the surface value and
depth by $\delta_{\gamma}$. At every fractional order the profile leaves the
exponential core and crosses onto a power law, with the classical profile
alone falling exponentially and departing from the family by many orders of
magnitude beyond $\zeta/\delta_{\gamma} \approx 10$. Compensating by
$(\zeta/\delta_{\gamma})^{1+\gamma}$ collapses each fractional profile onto
a plateau coinciding with the closed-form amplitude
$\gamma|A_{\gamma}|/(f\rho)$ of Eq.~(\ref{eq:tail}) with no fitted constant.
At $\gamma = 0.2$, $0.4$, $0.6$, and $0.8$ the measured
plateaus at $\zeta = 10^{5}$ match the closed-form values $0.172$, $0.269$,
$0.270$, and $0.174$ to relative errors between $3.10\times10^{-6}$ and
$8.39\times10^{-6}$. The amplitude is not monotone in $\gamma$: it rises to
a maximum of $0.282$ at $\gamma = 0.5$ and falls away on both sides, so
the pairs $\gamma = 0.2$ with $\gamma = 0.8$ and $\gamma = 0.4$ with
$\gamma = 0.6$ nearly coincide. That near-coincidence is a property of
$1/\Gamma(1-\gamma)$ and of the factor $\gamma$ multiplying it, and not a
plotting artifact.

\begin{figure}[H]
    \centering
    \includegraphics[width=\linewidth]{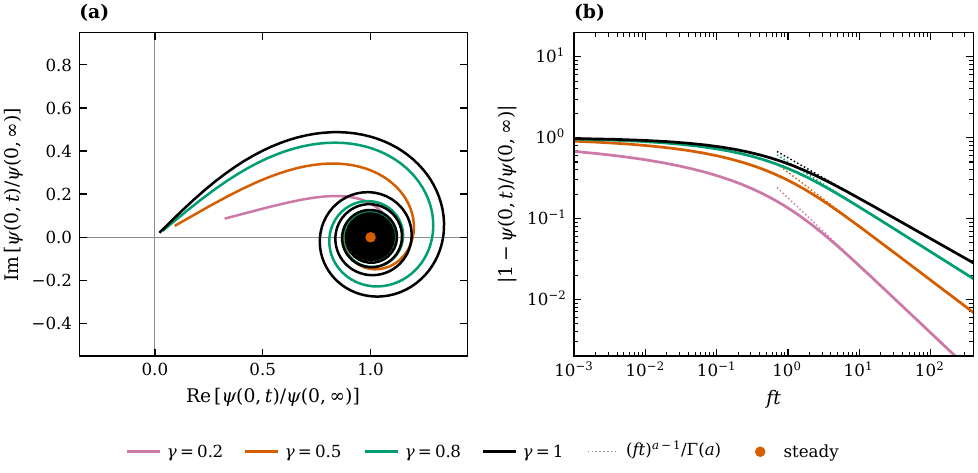}
    \caption{Surface response to a suddenly applied stress. (a) Trajectory
    of $\psi(0,t)/\psi(0,\infty)$ in the complex plane at four closure
    orders, converging on the steady state marked by the circle.
    (b) Magnitude of the departure from the steady state against $ft$, with
    the algebraic envelope $(ft)^{a-1}/\Gamma(a)$ of
    Eq.~(\ref{eq:envelope}) dotted. The decay is algebraic at every order,
    including $\gamma = 1$.}
    \label{fig:spinup}
\end{figure}

The response to a suddenly applied stress appears in
Fig.~\ref{fig:spinup}. The surface current approaches its steady value along
an inertially oscillating path whose amplitude decreases as the closure
becomes more nonlocal, since the exponent $a$ falls with $\gamma$. The
departure from the steady state falls onto the predicted envelope
$(ft)^{a-1}/\Gamma(a)$ at each order, with $a-1$ equal to $-0.833$,
$-0.667$, $-0.556$, and $-0.500$ at $\gamma = 0.2$, $0.5$, $0.8$, and
$1.0$. The decay is algebraic at every order including the classical one, so
no member of the family forgets its initial transient exponentially. The
closed form of Eq.~(\ref{eq:spinup}) agrees with the nested contour
inversion to between $2.97\times10^{-12}$ and $7.93\times10^{-12}$ for
$ft \leq 10$, and with the direct quadrature to between
$1.02\times10^{-15}$ and $4.70\times10^{-12}$ for $ft \leq 400$. The
classical case reproduces $\mathrm{erf}(\sqrt{ift})$ to
$1.64\times10^{-15}$.

\begin{figure}[H]
    \centering
    \includegraphics[width=\linewidth]{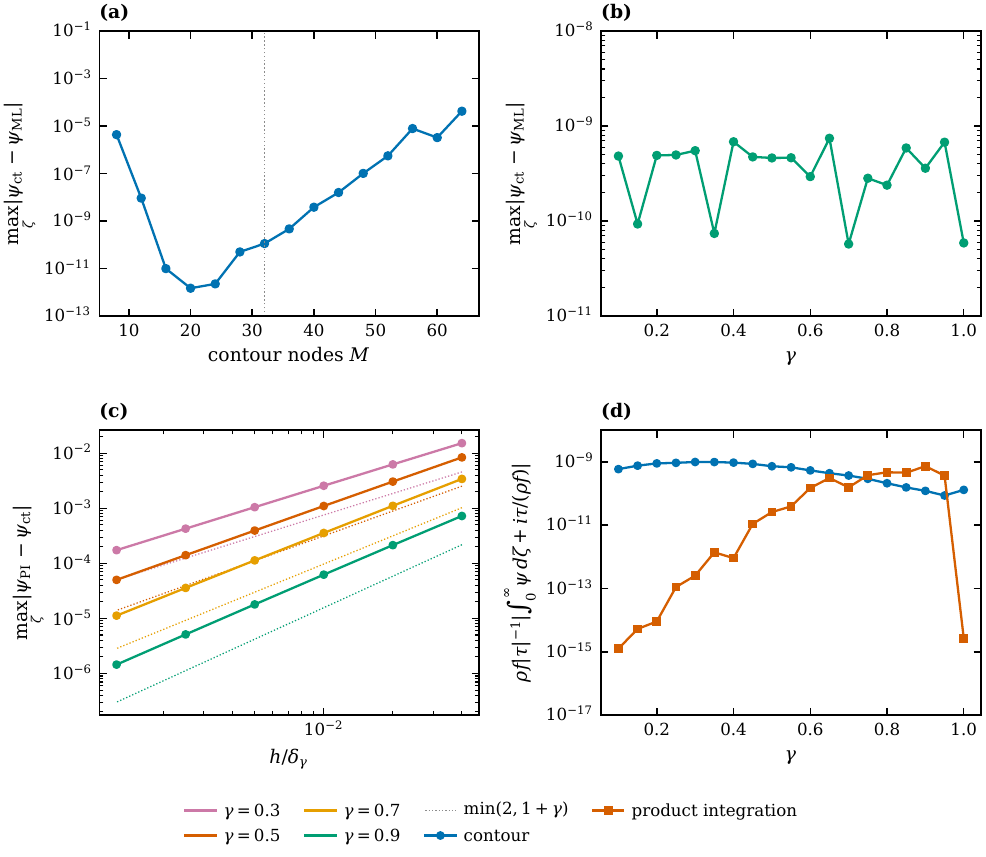}
    \caption{Numerical verification. (a) Maximum discrepancy between the
    contour inversion and the arbitrary precision Mittag-Leffler series
    against the number of contour nodes, with the working value $M = 32$
    marked. (b) The same discrepancy against closure order at $M = 32$.
    (c) Convergence of the product-integration march against grid spacing at
    four closure orders, with the predicted slopes $\min(2,1+\gamma)$
    dotted. (d) Relative residual of the transport invariant of
    Eq.~(\ref{eq:transport}) against closure order, evaluated along both the
    contour and the product-integration solutions.}
    \label{fig:verification}
\end{figure}

Verification is collected in Fig.~\ref{fig:verification} and
Table~\ref{tab:verification}. The contour rule reaches a residual near
$4\times10^{-12}$ at $M = 20$ to $24$ and degrades beyond it as roundoff is
amplified by $e^{2M/5}$; $M = 32$ is used throughout at a residual near
$10^{-10}$, chosen for uniform behaviour across the depth range rather than
for the best attainable accuracy at a single depth. Cross-method agreement
between the contour and the arbitrary precision series is between
$6\times10^{-11}$ and $8\times10^{-10}$ across $\gamma \in [0.1,1.0]$ with
no systematic trend in $\gamma$. Substituting the contour solution back into
the integral form of the closure, with the fractional integral evaluated by
arbitrary-precision quadrature, reproduces the velocity to between
$1.2\times10^{-10}$ and $5.0\times10^{-10}$ at $\gamma = 0.3$, $0.6$, and
$0.9$ and at depths of one and four scale depths. The
product-integration scheme converges at observed orders of $1.29$,
$1.48$, $1.66$, and $1.80$ at $\gamma = 0.3$, $0.5$, $0.7$, and $0.9$,
against the predicted $\min(2,1+\gamma)$ values of $1.3$, $1.5$, $1.7$, and
$1.9$. The shortfall grows with $\gamma$ and is consistent with a
preasymptotic regime set by the $\zeta^{\gamma}$ surface cusp rather than
with a formulation error. The transport invariant of
Eq.~(\ref{eq:transport}) is satisfied to $9.90\times10^{-10}$ along the
contour solution and to $7.29\times10^{-10}$ along the product-integration
solution, relative to $|\tau|/(\rho f)$. The product-integration residual
falls with decreasing $\gamma$ and collapses to $1.7\times10^{-15}$ at
$\gamma = 1$, because its flux-form update telescopes exactly in the local
limit, whereas the contour residual is set by the quadrature of a profile
carrying the surface cusp and is therefore nearly independent of $\gamma$.

\begin{table}[H]
\centering
\caption{Measured residuals of the verification protocol. All quantities are
dimensionless; transport residuals are relative to $|\tau|/(\rho f)$.}
\label{tab:verification}
\small
\begin{tabular}{lll}
\toprule
Check & Method pair & Residual \\
\midrule
Surface deflection, Eq.~(\ref{eq:theta}) & series vs closed form &
$1.07\times10^{-14}$ deg \\
Velocity, $0.25 < \zeta < 16$ & contour vs series &
$6.86\times10^{-10}$ \\
Closure, integral form of Eq.~(\ref{eq:volterra}) &
contour vs quadrature of $I^{\gamma}T$ & $5.0\times10^{-10}$ \\
Inversion pair for $E_{\mu,\nu}$ & contour vs series &
$1.4\times10^{-11}$ \\
Product quadrature, Eq.~(\ref{eq:piweights}) &
weights vs $I^{\gamma}\zeta^{2}$ at $N = 800$ & $2.7\times10^{-7}$ \\
Tail amplitude at $\zeta = 10^{5}$, Eq.~(\ref{eq:tail}) &
contour vs closed form & $3.22\times10^{-5}$ \\
Transport invariant, Eq.~(\ref{eq:transport}) & contour vs closed form &
$9.90\times10^{-10}$ \\
Transport invariant, Eq.~(\ref{eq:transport}) &
product integration vs closed form & $7.29\times10^{-10}$ \\
Response to applied stress, $ft \leq 10$ & closed form vs contour &
$8.57\times10^{-12}$ \\
Response to applied stress, $ft \leq 400$ & closed form vs quadrature &
$1.20\times10^{-11}$ \\
Local limit of the transient & closed form vs
$\mathrm{erf}(\sqrt{ift})$ & $1.64\times10^{-15}$ \\
\bottomrule
\end{tabular}
\end{table}

\section{Discussion}
\label{sec:discussion}

Equation~(\ref{eq:theta}) is the result of most immediate physical interest.
The surface deflection is a monotone function of the closure order alone,
independent of $K_{\gamma}$, $f$, $\rho$, and the magnitude of the wind
stress, and it is smaller than $45^{\circ}$ at every $\gamma < 1$. Inverting
it gives $\gamma \simeq 0.25$ for a deflection of $18^{\circ}$ and
$\gamma \simeq 0.5$ for $30^{\circ}$, which spans much of the range that
direct measurements report \cite{Price1999, Chereskin1995, Lenn2009,
Bressan2019}. That a single parameter reproduces the correct sign of the
discrepancy through a mechanism with no free amplitude is the principal
motivation for the construction. It does not amount to an explanation. No
calibration against observations or large-eddy simulation is attempted,
$K_{\gamma}$ is not derived from eddy statistics, and the deflection deficit
has at least three established alternative accounts, namely a
depth-dependent eddy viscosity \cite{Grisogono1995, Dritschel2020,
Constantin2021}, the diurnal cycling of near-surface stratification
\cite{Price1999}, and the contamination of estimates by geostrophic shear
\cite{Polton2013, Roach2015}. What the present result establishes is that
the deficit is also obtainable by weakening the locality of the
flux-gradient law while holding everything else fixed, and that this route
makes quantitative predictions elsewhere in the solution which the other
routes do not.

Two of those predictions are sharp enough to serve as discriminators. The
first is negative. Equation~(\ref{eq:transport}) is closure-free, so the
depth-integrated transport remains exactly normal to the wind at every
$\gamma$. A fractional closure therefore changes the surface angle without
touching the integrated transport, whereas explanations that modify the
vertical structure of the mixing generally alter both. Since the transport
is the better-observed of the two quantities \cite{Price1987,
Chereskin1995}, and since it is transport rather than surface velocity that
enters basin-scale circulation budgets and the time series used to diagnose
interbasin connectivity \cite{Herho2026itf}, the model survives its strongest
existing test by
construction, which is weak support but a useful constraint on any successor
model.

The second is positive and more vulnerable. The far field of
Eq.~(\ref{eq:tail}) is algebraic rather than exponential, with an exponent
of exactly $1+\gamma$ tied to the same order that fixes the surface angle. A
nonlocal closure of this type therefore predicts a wind-driven signal
persisting far below the nominal boundary layer, decaying as a power law
whose slope can be read directly off the observed surface deflection.
Observations of flattened spirals in which the rotation depth scale exceeds
the amplitude decay scale \cite{Price1999, Lenn2009} are qualitatively
consistent with a profile whose amplitude decays more slowly than the
classical exponential, but the correspondence is loose and the model makes a
considerably stronger claim than the available data can adjudicate. The
prediction is also where the model is most likely to fail physically, since
at depths of many Ekman scales the wind-driven signal is neither
horizontally homogeneous nor unstratified, and the assumption that the
memory of the closure extends over the entire overlying column ceases to be
defensible. A tempered kernel, cutting off the power law beyond a prescribed
range, would be the natural repair and would preserve the tractability of
the Laplace-transform solution.

The behaviour of the local limit deserves separate comment, since it is not
what one would expect of a one-parameter family containing the classical
case. The tail amplitude carries $1/\Gamma(1-\gamma)$, which vanishes
identically at $\gamma = 1$, so the algebraic far field does not shrink
continuously into the exponential one but is extinguished at the endpoint.
The consequence is visible in the winding number. For $\gamma$ below unity
the velocity vector makes a finite number of revolutions and then stops on
the direction fixed by Eq.~(\ref{eq:asymptotic}), while at $\gamma = 1$ it
winds without bound, and the transition between the two is effected by the
crossover depth of Eq.~(\ref{eq:crossover}) running off to infinity. The
logarithmic scaling of Eq.~(\ref{eq:windinglaw}) is offered as an account of
how that happens rather than as a proven result, and the agreement in
Table~\ref{tab:winding} is consistent with it over the range tested without
establishing it. Two features of the account survive independently of the
scaling law. The winding is carried by the pole at $p_{-1}$, which exists
only for $\gamma > 1/2$; and the entry of that pole onto the principal sheet
at $\gamma = 1/2$ produces no observable signature, because at that order
the algebraic tail still dominates the pole at every depth. The crossing is
a necessary but not a sufficient condition for extra revolutions, which
accounts for a null result that would otherwise be puzzling.

Placed against the wider literature on nonlocal closures, the present
construction is best read as an analytically tractable member of a family
otherwise studied numerically. Exact Green's-function expressions show the
eddy diffusivity to be an integral operator with finite support in space and
time, and direct numerical simulation has begun to characterize its shape
\cite{Hamba2022, Hamba2023, Hamba2025}. Macroscopic forcing methods have
produced systematic finite-rank approximations to the same operator
\cite{ManiPark2021, Shirian2022, Liu2023}. Where the measured kernel is
close to a power law, the operator is a fractional derivative and the
present solution applies with $\gamma$ read off the kernel exponent; where
it is not, the solution does not apply, and the discrepancy is informative.
The same logic underlies fractional Reynolds-stress and subgrid closures in
wall-bounded and scalar turbulence, where the fractional order has been
found to vary with distance from the wall rather than to take a single value
\cite{Song2021, AkhavanSafaei2023, Egolf2017}. A variable-order
generalization of Eq.~(\ref{eq:closure}) would be the corresponding
extension here, at the cost of the Laplace-transform solution and hence of
every closed form reported above. Boundary-layer schemes in wide operational
use already concede nonlocality by adding a countergradient term to a local
diffusivity \cite{Large1994, Stull1984}; the fractional closure differs in
representing the nonlocality within the operator rather than as an additive
correction to a local one, which is what makes the surface angle a function
of the closure order.

Several limitations bound the reach of everything above. The problem is
linear, steady or suddenly forced, horizontally homogeneous, unstratified,
and free of surface waves, so it omits the Stokes-Coriolis forcing, the
frontal and diurnal effects, and the superinertial response that recent
observations have shown to matter, including the leftward deflection
documented in the Bay of Bengal \cite{Polton2005, McWilliams2012,
Wenegrat2016, McPhaden2024}. The Caputo derivative is based at the surface,
so the stress at depth depends on the shear above it and not below it; that
choice follows from the direction of momentum input rather than from a
derivation, and a symmetric Riesz closure would forfeit the transform
solution and change the surface condition, with consequences not addressed
here. The coefficient $K_{\gamma}$ is a constant of the closure with
$\gamma$-dependent units, so absolute depth scales and speeds are not
comparable across the family and none is reported. Finally, the winding
number is measured rather than derived, its thresholds are resolved only to
the grid of closure orders used, and the double precision evaluation of the
deep turning fails beyond $\zeta \approx 10^{5}$ as $\gamma \to 1$, which is
why the identity was reverified at high precision.

\section{Conclusions}
\label{sec:conclusion}

The turbulent stress in a wind-driven boundary layer is exactly an integral
functional of the mean shear, and the local flux-gradient law is only the
leading term of an expansion presuming a mixing length short compared with
the scale over which the shear varies, a presumption that fails in a surface
layer whose energy-containing eddies span its own depth; requiring the memory
kernel to carry no preferred vertical scale then leaves a power law as the
sole admissible form, and the closure becomes a fractional derivative whose
order is the single free parameter of the theory. Imposed on the rotating
momentum balance, that closure cannot be applied to the velocity, since the
Caputo derivative of a bounded profile vanishes at the surface while its
Riemann-Liouville counterpart diverges there, leaving the wind stress
impossible to impose in the first case and the surface current unbounded in
the second; posed on the stress instead, the problem admits a closed solution
in Mittag-Leffler functions at every admissible order. The surface deflection
then depends on the closure order alone and is smaller than the classical
value throughout, reproducing the sign of the long-standing discrepancy
between theory and direct measurement through a mechanism carrying no free
amplitude, while the depth-integrated transport stays exactly normal to the
wind because that constraint survives any closure. The far field is algebraic
rather than exponential, with an exponent tied to the order that fixes the
deflection and an amplitude vanishing in the local limit, so that limit is
singular; the deep flow direction lies a quarter turn from the surface
direction modulo a full revolution at every fractional order, while the
number of revolutions taken to reach it grows without bound as the closure
becomes local; and the surface transient decays algebraically under a
suddenly applied stress, so no member of the family forgets its initial
condition exponentially. Whether such a closure survives contact with
stratification, surface waves, and the horizontally inhomogeneous flows in
which Ekman balances are measured would determine how far these conclusions
travel, and the natural extensions are a tempered kernel, bounding the reach
of the memory without destroying the transform solution, and a variable-order
closure, letting the degree of nonlocality vary with depth as measurements of
nonlocal eddy diffusivities suggest it should.

\section*{Declaration of competing interest}
The authors declare that they have no known competing financial interests or
personal relationships that could have appeared to influence the work
reported in this article.

\section*{Declaration of generative AI use}
During the preparation of this work, the authors used Claude Sonnet 5 solely
for the purposes of English grammar, vocabulary refinement, and improving
the overall readability of the manuscript. The authors maintain full
responsibility for the conceptualization, model development, mathematical
derivations, computational execution, and the analysis and interpretation
presented in this study.

\section*{Code and data availability}
This study uses no external data; the analysis code and every output it
produces, comprising fifteen CSV tables of per-panel figure data, the seven
figures in vector and raster form, and five plain-text computation reports,
are openly available at
\url{https://github.com/sandyherho/fractional_ekman_nonlocal}, where
\texttt{scripts/run\_all.py} regenerates all of it from a clean checkout.
The repository and its outputs are permanently archived on Zenodo: \url{https://doi.org/10.5281/zenodo.22830988}, which resolves to
the latest archived version and lists all prior ones, and both code and
archive are released under the MIT License.

\section*{Funding}
This study was supported by the Faculty of Earth Sciences and Technology
(FITB), Bandung Institute of Technology (ITB), through the PPMI Research
Program 2026 under Project ID FITB.PPMI-1-19-2026.


\begin{thebibliography}{99}

\bibitem{AbateValko2004} Abate, J.; Valk\'{o}, P. P. Multi-precision
Laplace transform inversion. \textit{Int. J. Numer. Methods Eng.}
\textbf{2004}, \textit{60}, 979--993.
\url{https://doi.org/10.1002/nme.995}

\bibitem{AkhavanSafaei2023} Akhavan-Safaei, A.; Zayernouri, M. A non-local
spectral transfer model and new scaling law for scalar turbulence.
\textit{J. Fluid Mech.} \textbf{2023}, \textit{956}, A26.
\url{https://doi.org/10.1017/jfm.2022.1066}

\bibitem{Berkowicz1980} Berkowicz, R.; Prahm, L. P. On the spectral
turbulent diffusivity theory for homogeneous turbulence. \textit{J. Fluid
Mech.} \textbf{1980}, \textit{100(2)}, 433--448. \url{https://doi.org/10.1017/S0022112080001231}

\bibitem{Bressan2019} Bressan, A.; Constantin, A. The deflection angle of
surface ocean currents from the wind direction. \textit{J. Geophys. Res.
Oceans} \textbf{2019}, \textit{124}. 7412–-7420. 
\url{https://doi.org/10.1029/2019JC015454}

\bibitem{Caputo1967} Caputo, M. Linear Models of Dissipation whose Q is almost Frequency Independent—-II. \textit{Geophys. J. Int.}
\textbf{1967}, \textit{13(5)}, 529--539.
\url{https://doi.org/10.1111/j.1365-246X.1967.tb02303.x}

\bibitem{Chereskin1995} Chereskin, T. K. Direct evidence for an Ekman
balance in the California Current. \textit{J. Geophys. Res. Oceans}
\textbf{1995}, \textit{100}(C9), 18261--18269.
\url{https://doi.org/10.1029/95JC02182}

\bibitem{Constantin2021} Constantin, A. Frictional effects in wind-driven
ocean currents. \textit{Geophys. Astrophys. Fluid Dyn.} \textbf{2021},
\textit{115(1)}, 1--14.
\url{https://doi.org/10.1080/03091929.2020.1748614}

\bibitem{Corrsin1974} Corrsin, S. Limitations of gradient transport models
in random walks and in turbulence. \textit{Adv. Geophys.} \textbf{1974},
\textit{18A}, 25--60.

\bibitem{Diethelm2010} Diethelm, K. \textit{The Analysis of Fractional
Differential Equations: An Application-Oriented Exposition Using
Differential Operators of Caputo Type}; Lecture Notes in Mathematics;
Springer: Berlin, Germany, 2010.
\url{https://doi.org/10.1007/978-3-642-14574-2}

\bibitem{DiethelmFordFreed2002} Diethelm, K.; Ford, N. J.; Freed, A. D. A Predictor-Corrector Approach for the Numerical Solution of Fractional Differential Equations. \textit{Nonlinear Dyn.} \textbf{2002}, \textit{29},
3--22.
\url{https://doi.org/10.1023/A:1016592219341}

\bibitem{DiethelmFordFreed2004} Diethelm, K.; Ford, N. J.; Freed, A. D.
Detailed Error Analysis for a Fractional Adams Method. \textit{Numer.
Algorithms} \textbf{2004}, \textit{36}, 31--52.
\url{https://doi.org/10.1023/B:NUMA.0000027736.85078.be}

\bibitem{DiethelmFordFreedLuchko2005} Diethelm, K.; Ford, N. J.; Freed,
A. D.; Luchko, Y. Algorithms for the fractional calculus: A selection of
numerical methods. \textit{Comput. Methods Appl. Mech. Eng.} \textbf{2005},
\textit{194(6--8)}, 743--773.
\url{https://doi.org/10.1016/j.cma.2004.06.006}

\bibitem{Dixon1985} Dixon, J. On the order of the error in discretization
methods for weakly singular second kind Volterra integral equations with
non-smooth solutions. \textit{BIT} \textbf{1985}, \textit{25}, 624--634.

\bibitem{Dritschel2020} Dritschel, D. G.; Paldor, N.; Constantin, A. The
Ekman spiral for piecewise-uniform viscosity. \textit{Ocean Sci.}
\textbf{2020}, \textit{16(5)}, 1089--1093.
\url{https://doi.org/10.5194/os-16-1089-2020}

\bibitem{Egolf2017} Egolf, P. W.; Hutter, K. Fractional turbulence models.
In \textit{Progress in Turbulence VII}; Springer Proceedings in Physics 196;
Springer: Cham, Switzerland, 2017.
\url{https://doi.org/10.1007/978-3-319-57934-4_18}

\bibitem{Ekman1905} Ekman, V. W. On the influence of the Earth's rotation on
ocean currents. \textit{Ark. Mat. Astron. Fys.} \textbf{1905}, \textit{2},
1--53.

\bibitem{Elipot2009} Elipot, S.; Gille, S. T. Ekman layers in the Southern
Ocean: spectral models and observations, vertical viscosity and boundary
layer depth. \textit{Ocean Sci.} \textbf{2009}, \textit{5(2)}, 115--139.
\url{https://doi.org/10.5194/os-5-115-2009}

\bibitem{Garrappa2015} Garrappa, R. Numerical Evaluation of Two and Three Parameter Mittag-Leffler Functions. \textit{SIAM J. Numer. Anal.}
\textbf{2015}, \textit{53(3)}, 1350--1369.
\url{https://doi.org/10.1137/140971191}

\bibitem{Gorenflo2002} Gorenflo, R.; Loutchko, J.; Luchko, Y. Computation of
the Mittag-Leffler function $E_{\alpha,\beta}(z)$ and its derivative.
\textit{Fract. Calc. Appl. Anal.} \textbf{2002}, \textit{5(4)}.

\bibitem{Gorenflo2014} Gorenflo, R.; Kilbas, A. A.; Mainardi, F.; Rogosin,
S. V. \textit{Mittag-Leffler Functions, Related Topics and Applications};
Springer: Berlin, Germany, 2014.
\url{https://doi.org/10.1007/978-3-662-43930-2}

\bibitem{Grisogono1995} Grisogono, B. A generalized Ekman layer profile with
gradually varying eddy diffusivities. \textit{Q. J. R. Meteorol. Soc.}
\textbf{1995}, \textit{121}, 445--453. \url{https://doi.org/10.1002/qj.49712152211}

\bibitem{Hamba2022} Hamba, F. Analysis and modelling of non-local eddy
diffusivity for turbulent scalar flux. \textit{J. Fluid Mech.}
\textbf{2022}, \textit{950}, A38.
\url{https://doi.org/10.1017/jfm.2022.842}

\bibitem{Hamba2023} Hamba, F. Non-local eddy diffusivity model based on
turbulent energy density in scale space. \textit{J. Fluid Mech.}
\textbf{2023}, \textit{977}, A11.
\url{https://doi.org/10.1017/jfm.2023.969}

\bibitem{Hamba2025} Hamba, F. Analysis and modelling of non-local eddy
diffusivity in turbulent channel flow. \textit{J. Fluid Mech.}
\textbf{2025}, \textit{1012}, A21.
\url{https://doi.org/10.1017/jfm.2025.10221}

\bibitem{Harris2020} Harris, C. R.; Millman, K. J.; van der Walt, S. J.;
Gommers, R.; Virtanen, P.; Cournapeau, D.; Wieser, E.; Taylor, J.;
Berg, S.; Smith, N. J.; Kern, R.; Picus, M.; Hoyer, S.; van Kerkwijk, M. H.;
Brett, M.; Haldane, A.; del R{\'\i}o, J. F.; Wiebe, M.; Peterson, P.;
G{\'e}rard-Marchant, P.; Sheppard, K.; Reddy, T.; Weckesser, W.;
Abbasi, H.; Gohlke, C.; Oliphant, T. E. Array programming with NumPy.
\textit{Nature} \textbf{2020}, \textit{585}, 357--362.
\url{https://doi.org/10.1038/s41586-020-2649-2}

\bibitem{Herho2025kh2d} Herho, S. H. S.; Trilaksono, N. J.; Fajary, F. R.;
Napitupulu, G.; Anwar, I. P.; Khadami, F.; Irawan, D. E. kh2d-solver: A
Python library for idealized two-dimensional incompressible Kelvin-Helmholtz
instability. \textit{Appl. Comput. Mech.} \textbf{2025}, \textit{19},
125--156.
\url{https://doi.org/10.24132/acm.2025.1040}

\bibitem{Herho2026itf} Herho, S. H. S.; Herho, K. E. P.; Anwar, I. P.;
Cahyarini, S. Y. Strengthening ITF and Weakening AMOC: Time Series Evidence of Trends and Causal Pathways to Agulhas Variability. \textit{Ocean Coast.
Res.} \textbf{2026}, \textit{74}, e26002.
\url{https://doi.org/10.1590/2675-2824074.25092}

\bibitem{Herho2026wave} Herho, S. H. S.; Anwar, I. P.; Khadami, F.; Ndruru,
T. R. E. B. N.; Suwarman, R.; Irawan, D. E. wave-attenuation-1d: An idealized
one-dimensional framework for wave attenuation through coastal vegetation
using Numba-accelerated shallow water equations. \textit{J. Theor. Appl.
Mech.} \textbf{2026}, \textit{56}, 89--102.
\url{https://doi.org/10.55787/jtams.2026.1.AI00236}

\bibitem{Hunter2007} Hunter, J. D. Matplotlib: A 2D graphics environment.
\textit{Comput. Sci. Eng.} \textbf{2007}, \textit{9(3)}, 90--95.
\url{https://doi.org/10.1109/MCSE.2007.55}

\bibitem{Irawan2026amerta} Irawan, D. E.; Herho, S. H. S.; Anwar, I. P.;
Khadami, F.; Pamumpuni, A.; Kartiko, R. D.; Riawan, E.; Suwarman, R.;
Puradimaja, D. J. amerta: A Python library for idealized 1D Saint-Venant
dam-break simulation. \textit{Front. Water} \textbf{2026}, \textit{8},
1900409.
\url{https://doi.org/10.3389/frwa.2026.1900409}

\bibitem{Irawan2026kdv} Irawan, D. E.; Herho, S. H. S.; Pamumpuni, A.;
Kartiko, R. D.; Khadami, F.; Anwar, I. P.; Sujatmiko, K. A.; Handayani,
A. P.; Fajary, F. R.; Suwarman, R. An Open-Source Pseudo-Spectral Solver for Idealized Korteweg–de Vries Soliton Simulations. \textit{Water}
\textbf{2026}, \textit{18(7)}, 779.
\url{https://doi.org/10.3390/w18070779}

\bibitem{Large1994} Large, W. G.; McWilliams, J. C.; Doney, S. C. Oceanic
vertical mixing: A review and a model with a nonlocal boundary layer
parameterization. \textit{Rev. Geophys.} \textbf{1994}, \textit{32(4)},
363--403.
\url{https://doi.org/10.1029/94RG01872}

\bibitem{Lenn2009} Lenn, Y.-D.; Chereskin, T. K. Observations of Ekman
currents in the Southern Ocean. \textit{J. Phys. Oceanogr.} \textbf{2009},
\textit{39}, 768--779.
\url{https://doi.org/10.1175/2008JPO3943.1}

\bibitem{Liu2023} Liu, J.; Williams, H. H.; Mani, A. Systematic approach for
modeling a nonlocal eddy diffusivity. \textit{Phys. Rev. Fluids}
\textbf{2023}, \textit{8}, 124501.
\url{https://doi.org/10.1103/PhysRevFluids.8.124501}

\bibitem{Lubich1985} Lubich, C. Fractional linear multistep methods for
Abel-Volterra integral equations of the second kind. \textit{Math. Comput.}
\textbf{1985}, \textit{45}, 463--469.
\url{https://doi.org/10.1090/S0025-5718-1985-0804935-7}

\bibitem{Lubich1986} Lubich, C. Discretized Fractional Calculus.
\textit{SIAM J. Math. Anal.} \textbf{1986}, \textit{17(3)}, 704--719.
\url{https://doi.org/10.1137/0517050}

\bibitem{Mainardi2010} Mainardi, F. \textit{Fractional Calculus and Waves in
Linear Viscoelasticity}; Imperial College Press: London, UK, 2010.
\url{https://doi.org/10.1142/p614}

\bibitem{ManiPark2021} Mani, A.; Park, D. Macroscopic forcing method: A tool
for turbulence modeling and analysis of closures. \textit{Phys. Rev. Fluids}
\textbf{2021}, \textit{6}, 054607.
\url{https://doi.org/10.1103/PhysRevFluids.6.054607}

\bibitem{McPhaden2024} McPhaden, M. J.; Athulya, K.; Girishkumar, M. S.;
Orli\'{c}, M. Ekman revisited: Surface currents to the left of the winds in
the Northern Hemisphere. \textit{Sci. Adv.} \textbf{2024}, \textit{10(46)},
eadr0282.
\url{https://doi.org/10.1126/sciadv.adr0282}

\bibitem{McWilliams2012} McWilliams, J. C.; Huckle, E.; Liang, J.-H.;
Sullivan, P. P. The Wavy Ekman Layer: Langmuir Circulations, Breaking Waves, and Reynolds Stress. \textit{J. Phys. Oceanogr.} \textbf{2012}, \textit{42},
1793--1816. \url{https://doi.org/10.1175/JPO-D-12-07.1}

\bibitem{MetzlerKlafter2000} Metzler, R.; Klafter, J. The random walk's
guide to anomalous diffusion: a fractional dynamics approach.
\textit{Phys. Rep.} \textbf{2000}, \textit{339(1)}, 1--77.
\url{https://doi.org/10.1016/S0370-1573(00)00070-3}

\bibitem{Podlubny1999} Podlubny, I. \textit{Fractional Differential
Equations}; Mathematics in Science and Engineering 198; Academic Press: San
Diego, CA, 1999.

\bibitem{Polton2005} Polton, J. A.; Lewis, D. M.; Belcher, S. E. The Role of Wave-Induced Coriolis–Stokes Forcing on the Wind-Driven Mixed Layer.
\textit{J. Phys. Oceanogr.} \textbf{2005}, \textit{35}, 444--457. \url{https://doi.org/10.1175/JPO2701.1}.

\bibitem{Polton2013} Polton, J. A.; Lenn, Y.-D.; Elipot, S.; Chereskin,
T. K.; Sprintall, J. Can Drake Passage Observations Match Ekman's Classic Theory?
 \textit{J. Phys. Oceanogr.} \textbf{2013}, \textit{43}, 1733--1740.
\url{https://doi.org/10.1175/JPO-D-13-034.1}

\bibitem{Price1987} Price, J. F.; Weller, R. A.; Schudlich, R. R.
Wind-Driven Ocean Currents and Ekman Transport. \textit{Science}
\textbf{1987}, \textit{238(4833)}, 1534--1538.
\url{https://doi.org/10.1126/science.238.4833.1534}

\bibitem{Price1999} Price, J. F.; Sundermeyer, M. A. Stratified Ekman
layers. \textit{J. Geophys. Res. Oceans} \textbf{1999}, \textit{104}(C9),
20467--20494.
\url{https://doi.org/10.1029/1999JC900164}

\bibitem{Roach2015} Roach, C. J.; Phillips, H. E.; Bindoff, N. L.; Rintoul,
S. R. Detecting and Characterizing Ekman Currents in the Southern Ocean.
\textit{J. Phys. Oceanogr.} \textbf{2015}, \textit{45}, 1205--1223.
\url{https://doi.org/10.1175/JPO-D-14-0115.1}

\bibitem{ScottBlair1947} Scott Blair, G. W.; Veinoglou, B. C.; Caffyn, J. E.
Limitations of the Newtonian time scale in relation to non-equilibrium
rheological states and a theory of quasi-properties. \textit{Proc. R. Soc.
London Ser. A} \textbf{1947}, \textit{189(1016)}, 69--87.
\url{https://doi.org/10.1098/rspa.1947.0029}

\bibitem{ScottBlair1949} Scott Blair, G. W.; Caffyn, J. E. VI. An application of the theory of quasi-properties to the treatment of anomalous strain-stress relations. \textit{London Edinburgh Dublin Philos. Mag. J. Sci.}
\textbf{1949}, \textit{40(300)}, 80--94.
\url{https://doi.org/10.1080/14786444908561213}

\bibitem{Shirian2022} Shirian, Y.; Mani, A. Eddy diffusivity operator in
homogeneous isotropic turbulence. \textit{Phys. Rev. Fluids} \textbf{2022},
\textit{7}, L052601.
\url{https://doi.org/10.1103/PhysRevFluids.7.L052601}

\bibitem{Song2021} Song, F.; Karniadakis, G. E. Variable-Order Fractional Models for Wall-Bounded Turbulent Flows. \textit{Entropy} \textbf{2021},
\textit{23(6)}, 782.
\url{https://doi.org/10.3390/e23060782}

\bibitem{Stull1984} Stull, R. B. Transilient Turbulence Theory. Part I: The
Concept of Eddy Mixing across Finite Distances. \textit{J. Atmos. Sci.}
\textbf{1984}, \textit{41}, 3351--3367. \url{https://doi.org/10.1175/1520-0469(1984)041%3C3351:TTTPIT%3E2.0.CO;2}

\bibitem{Talbot1979} Talbot, A. The Accurate Numerical Inversion of Laplace Transforms. \textit{IMA J. Appl. Math.} \textbf{1979}, \textit{23(1)}, 97--120.
\url{https://doi.org/10.1093/imamat/23.1.97}

\bibitem{Virtanen2020} Virtanen, P.; Gommers, R.; Oliphant, T. E.;
Haberland, M.; Reddy, T.; Cournapeau, D.; Burovski, E.; Peterson, P.;
Weckesser, W.; Bright, J.; van der Walt, S. J.; Brett, M.; Wilson, J.;
Millman, K. J.; Mayorov, N.; Nelson, A. R. J.; Jones, E.; Kern, R.;
Larson, E.; Carey, C. J.; Polat, {\.I}.; Feng, Y.; Moore, E. W.;
VanderPlas, J.; SciPy 1.0 Contributors. SciPy 1.0: Fundamental
algorithms for scientific computing in Python. \textit{Nat. Methods} \textbf{2020},
\textit{17}, 261--272.
\url{https://doi.org/10.1038/s41592-019-0686-2}

\bibitem{Wenegrat2016} Wenegrat, J. O.; McPhaden, M. J. Wind, Waves, and Fronts: Frictional Effects in a Generalized Ekman Model. \textit{J. Phys.
Oceanogr.} \textbf{2016}, \textit{46}, 371--394.
\url{https://doi.org/10.1175/JPO-D-15-0162.1}

\bibitem{Wong2011} Wong, B. Points of view: Color blindness.
\textit{Nat. Methods} \textbf{2011}, \textit{8}, 441.
\url{https://doi.org/10.1038/nmeth.1618}

\end{thebibliography}
\end{document}